MATHEMATICAL MODEL OF CONCENTRATING SOLAR COOKER

González Avilés, Mauricio[1], González-Avilés, José Juan[2]

[1]*Universidad Michoacana de San Nicolás de Hidalgo, Avenida Francisco J. Mujica S/N ciudad Universitaria. C:P. 58030, Morelia, Michoacán-México*
[2]Universidad Intercultural Indígena de Michoacán, Carretera Pátzcuaro-Huecorio s/n, Pátzcuaro, Michoacán C.P.61614.

Abstract
The main purpose of this work is to obtain a mathematical model consistent with the thermal behavior of concentrating solar cookers, such as "Jorhejpataranskua". We also want to simulate different conditions respect to the parameters involved of several materials for its construction and efficiency. The model is expressed in terms of a coupled nonlinear system of differential equations which are solved using Mathematica 8. The results obtained by our model are compared with measurements of solar cooker in field testing operation. We obtained good results in agreement with experimental data. Moreover, the simulation results are used by calculating cooking power and standardized cooking power of solar cooker for different parameters.

INTRODUCTION
Development of solar thermal technologies, particularly solar cookers, where it is necessary the application of testing protocols, for which is required instrumentation and equipment, although many times is not easy to get them. Furthermore when weather conditions are not favorable we cannot make testing protocols. Therefore, we propose a theoretical thermal model based on analysis of thermal physics of solar cooker "Jorhejpatarnskua" [1], [2], [3], which is a solar cooker based on a Compound Parabolic Concentrator of revolution (CPC) with an aluminum cylindrical container as absorber, so this is use to simulate the water temperature and estimate cooking power, for this purpose we make an experimental design due to simulate solar radiation with an arrangement of lamps, similar to the one considered in [4]. The paper is organized as follows:

- Thermal model description
- Analysis and equations of heat transfer
- Description of experimental design
- Results and comparison with field testing

**Nomenclature**

$Q_{1rad}$ Radiation heat flux directly incident over the absorber container
$Q_{2rad}$ Radiation heat flux between container and sky
$Q_{3conv}$ Convective heat flux between container and environment
$Q_{3conv2}$ Convective heat flux between absorber container and air inside stove
$Q_{4rad}$ Radiation heat flux between container and reflectors
$Q_{5rad}$ Radiation heat flux directly incident over reflectors
$Q_{6conv}$ Convective heat flux between reflectors and environment
$Q_{7rad}$ Radiation heat flux between reflectors and sky
$Q_{8conv}$ Convective heat flux between container surface's and its inside
$Q_{9rad}$ Radiation heat flux from container to fluid

$m_r$ Mass of container

$m_{rf}$ Mass of reflector sheet

$m_f$ Mass of water

$c_r$ Specific heat of aluminum container

$c_{rf}$ Specific heat of aluminum reflectors

$c_f$ Specific heat of water

$T_r$ Mean temperature of container

$T_{rf}$ Mean surface temperature of reflector sheet

$T_f$ Mean water temperature

$T_{amb}$ Environment temperature

$T_{sky}$ Sky temperature

$t$ Time

$A_{rf}$ Solar colector area of solar cooker

$A_r$ Absorber container area

$\alpha$ Absorptance of absorber

$\alpha_{rf}$ Absorptance of reflectors

$\eta_0$ Thermal efficiency

$\rho_m$ Reflectance of sheet

$n$ Mean number of reflections inside CPC

$\varepsilon_r$ Emittance of absorber

$\varepsilon_{rf}$ Emittance of reflectors

$h_{r,amb}$ Convective heat coefficient between absorber and environment

$h_{r,int2}$ Convective heat coefficient between absorber and air inside solar cooker

$h_{rf,amb}$ Convective heat coefficient between reflectors and environment

$h_{r,inte}$ Convective heat coefficient between container and fluid

$I_D$ Direct irradiance

$I_R$ Reflected irradiance

$\sigma$ Stefan Boltzmann constant

$P_r$ Cooking power

$P_{s\tan dard}$ Standardized cooking power

$\Delta T$ Temperature difference between fluid and environment

DESCRIPTION OF THERMAL MODEL

The heat transfer model proposed in this paper is based on using energy balance over three components of solar cooker: absorber container, reflective sheets and fluid.

*Balance energy equations*

*Heat transfer equation of container*

We make an energy balance of heat transfer inside of absorber container of solar cooker, this is described by the equation:

$$m_r c_r \frac{dT_r}{dt} = Q_{1\text{rad}} - -Q_{2\text{rad}} - Q_{3\text{conv}} - Q_{4\text{rad}} - Q_{8\text{conv}} - Q_{9\text{rad}} - Q_{3\text{conv2}} \qquad (1)$$

Where

| | |
|---|---|
| $Q_{1\text{rad}} = A_r \alpha I_D + A_{rf} I_R$ | $Q_{2\text{rad}} = A_r \varepsilon_r \sigma \left(T_r^4 - T_{cielo}^4\right)$ |
| Stefan Boltzmann constant $\sigma = 5.669\, X\, 10^{-8} \frac{W}{m^2 C^4}$ | Emittance of absorber $\varepsilon_r = 0.5$ |
| $Q_{2\text{rad}}\ A_r \varepsilon_r \sigma \left(T_r^4 - T_{cielo}^4\right)$ | $T_{sky} = 0.0552\, T_{amb}^{1.5}$ |
| $Q_{3\text{conv}}\ A_r h_{r,amb} \left(T_r - T_{amb}\right)$ | $Q_{3\text{conv2}}\ A_r h_{r,int2} \left(T_r - T_{int2}\right)$ |
| $T_{int2} = \frac{T_r + T_{rf}}{2}$ | $Q_{4\text{rad}}\ A_r \varepsilon_r \sigma \left(T_r^4 - T_{rf}^4\right)$ |
| $Q_{8\text{conv}}\ A_r h_{r,inte} \left(T_{inte} - T_r\right)$ | $Q_{9\text{rad}}\ A_r \varepsilon_r \sigma \left(T_r^4 - T_f^4\right)$ |
| $T_{inte} = \frac{T_r + T_f}{2}$ | |

*Heat transfer equation of reflectors*

Apliying a balance energy for reflector sheets we obtain the next equation:

$$m_{rf} c_{rf} \frac{dT_{rf}}{dt} Q_{5\text{rad}} + Q_{4\text{rad}} - Q_{6\text{conv}} - Q_{7\text{rad}} + Q_{3\text{conv2}} \qquad (2)$$

Where

$$Q_{7\text{rad}}\ A_{rf} \varepsilon_{rf} \sigma \left(T_{rf}^4 - T_{sky}^4\right)$$

*Heat transfer equation of fluid*

Similarly to the past two equations, we obtain:

$$m_f c_f \frac{dT_f}{dt} Q_{8\mathrm{conv}} + Q_{9\mathrm{rad}} \qquad (3)$$

*Equations (1), (2) y (3), form a coupled system of nonlinear differential equations for variables* $T_f$ , $T_{rf}$ *and* $T_r$ .

SOLUTION OF THE SYSTEM OF NONLINEAR DIFFRENTIAL EQUATIONS

In order to solve numerically the system of equations (1), (2) and (3) we used the scientific mathematical program called Mathematica version 8. We defined full equations with their consistent physical units, i.e., for heat (Watts), temperature is measured in degrees (°C) and time in minutes (min).

To solve the equations (1), (2) and (3) it is necesary give initial conditions, in this case we used initial conditions as follows:

$$T_r(t=0) T_{amb} \quad T_{rf}(t=0) T_{amb} \quad T_{rf}(t=0) T_{amb} \ , \qquad (4)$$

where $T_{amb}$ is environment temperature.

Another initial conditions could be the values obtained by experimental testing.

Then we use the Mathematica comand called NDSolve, which gives numerical solutions of a system of differential equations of arbitrary order, as long as it has well defined initial conditions. In this case we used an automathic method, which is the best method avaliable for that kind of system chosen by Mathematica, generally is the higher order method (see Appendix A).

*Cooking power estimation*

By solving the system of equations we obtained the numerical values of mean water temperature, so that we can estimate cooking power by using right hand side of equation (3).

Standardized cooking power is calculated as a función of temperature difference between fluid temperature and environment temperature, i.e. ( $\Delta T = T_f - T_{amb}$ ). Then standardized cooking power appears when temperature difference becomes 50 degrees.

RESULTS AND CMPARISON WITH FIELD TESTING

In the next figures below we show simulation results (solid line) and experimental data (points) in the case of mean water temperature. We also show results for cooking power, and finally we show results for mean reflectors temperature and mean container temperature only for two cases: pressure aluminuim pot with aluminium reflector sheets and pressure aluminium pot with gift paper as reflector sheets.

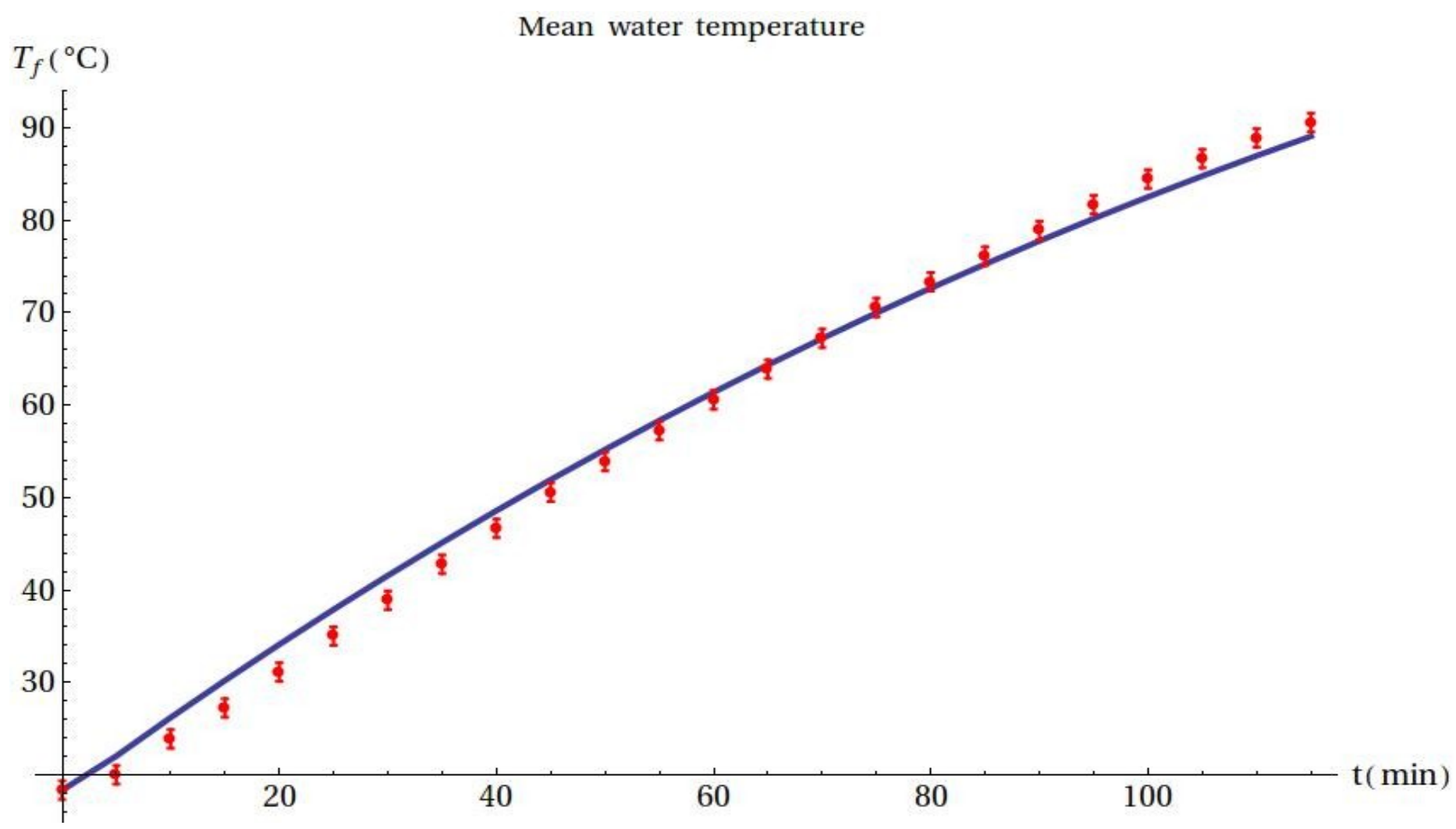

Figure 3.1: Show the mean water temperature for mass water of 4.2 kg versus time for an aluminium pressure pot. Numerical solution is represent by solid line and experimental data by dots with error bars. Mean water temperature increases in time up to boiling point in t=110 min aproximately. The relative error between simulation and experimental data is 2.89%.

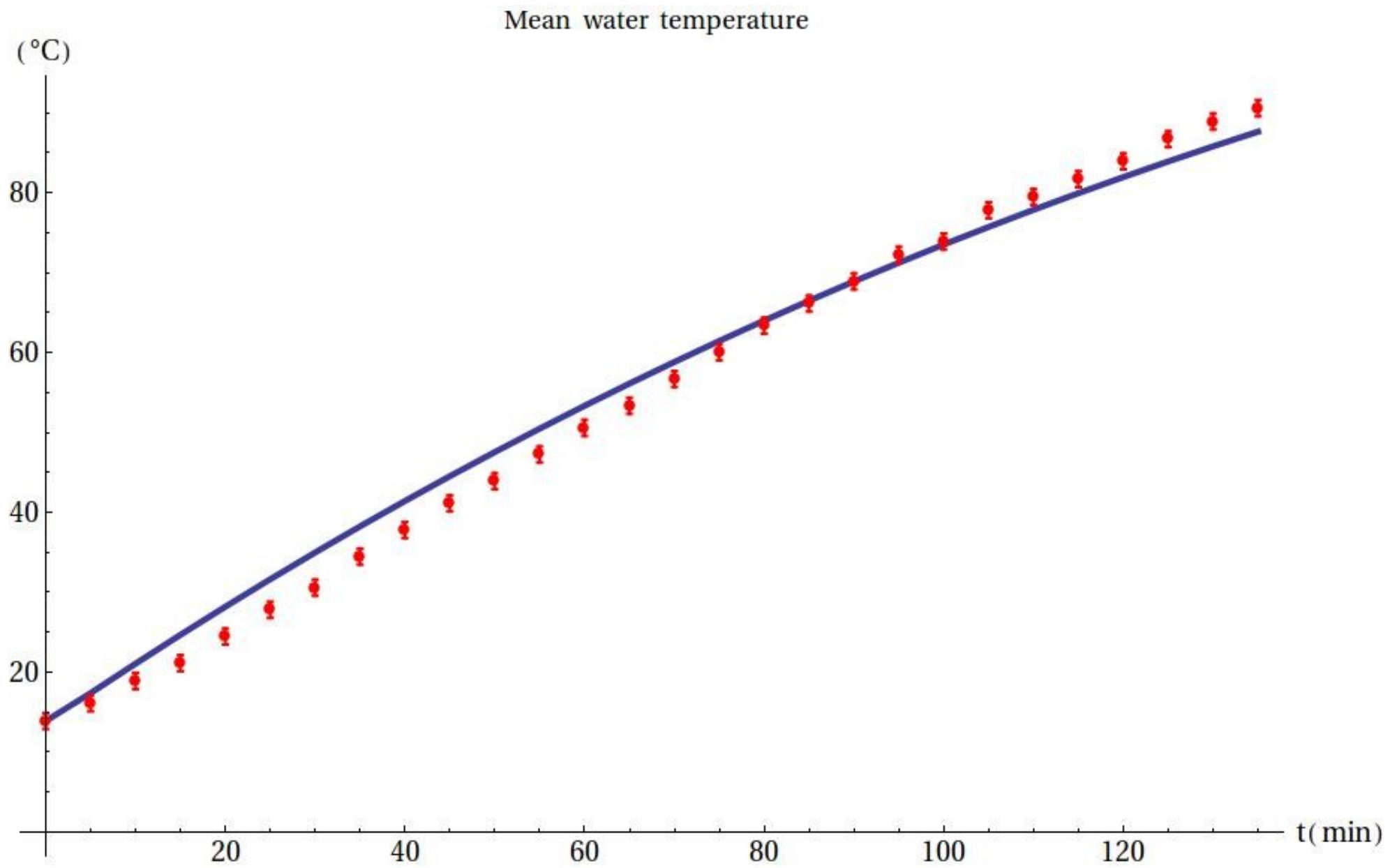

Figure 3.2: Show the mean water temperature for mass water of 4.2 kg versus time. In this case the container is made of stainless steel. Numerical solution is represent by solid line and experimental data by dots with error bars. Mean water temperature increases in time up to boiling point in t=135 min aproximately. The relative error between simulation and experimental data is 4.43%.

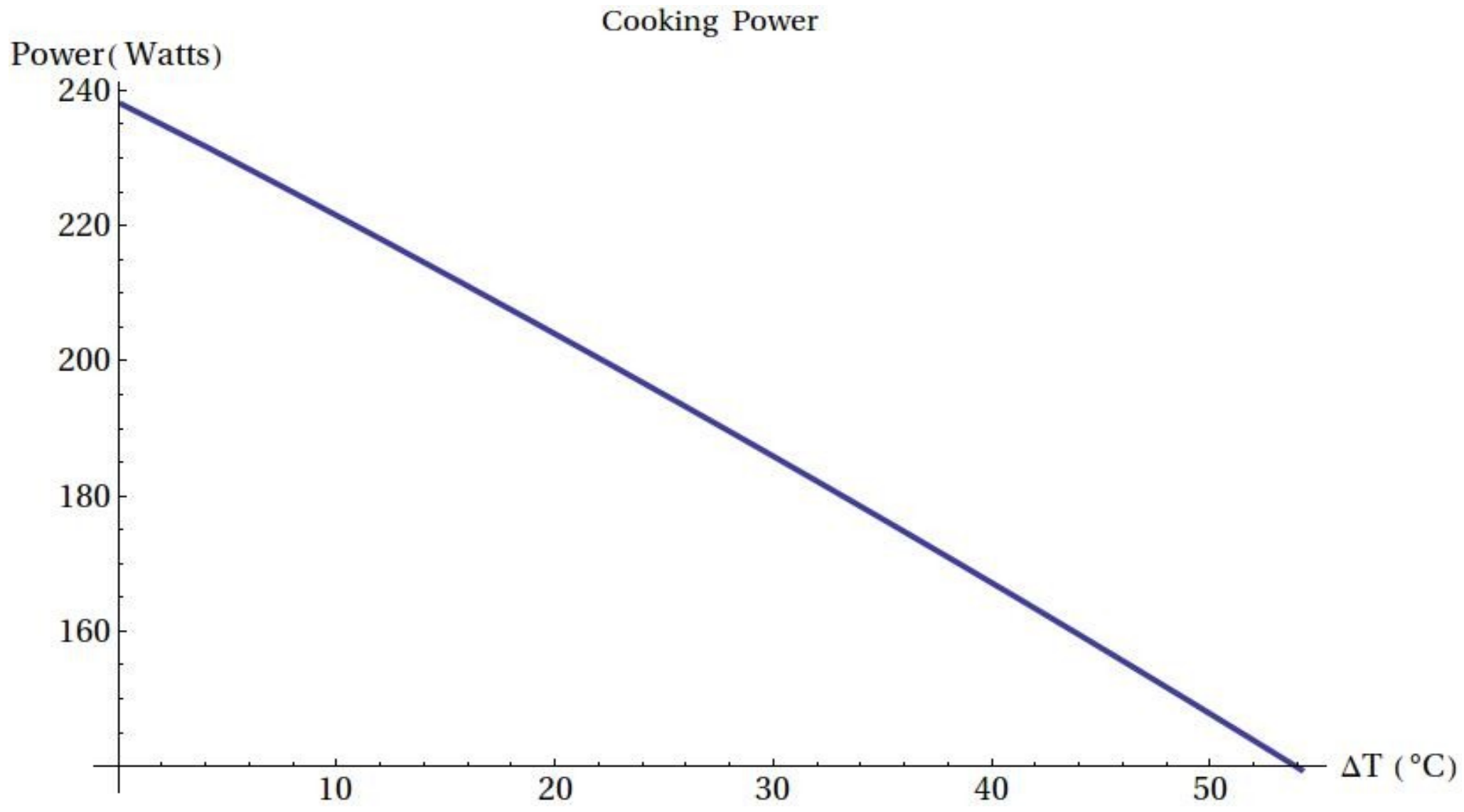

Figure 3.3: Show the cooking power versus temperature difference corresponding to parameters of Fig. (3.1). Standardized cooking power is equal to 112 Watts.

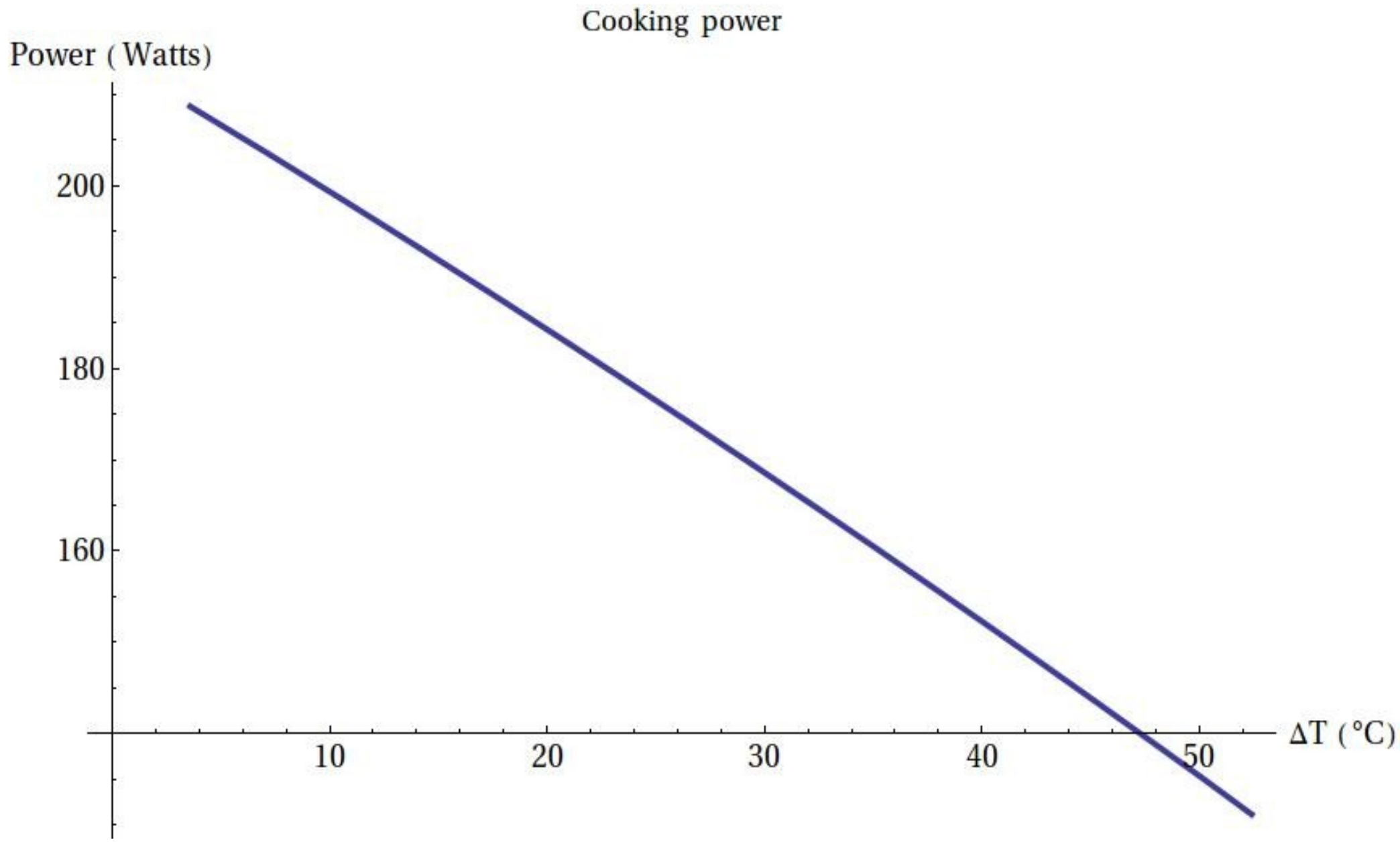

Figure 3.4: Show the cooking power versus temperature difference corresponding to parameters of Fig. (3.2). Standardized cooking power is equal to 104.4 Watts.

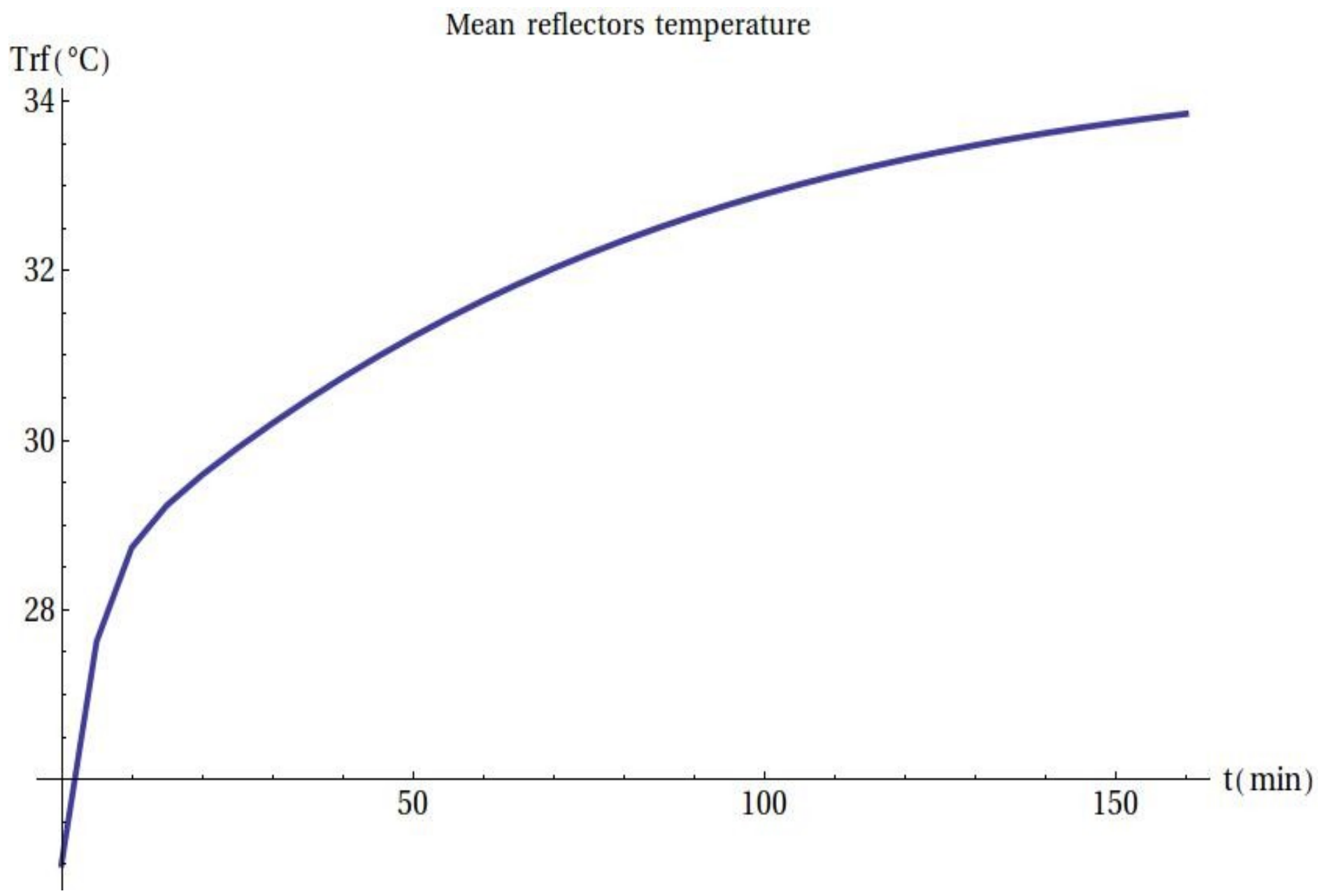

Figure 3.5: Show the mean reflectors temperature versus time corresponding to pressure aluminum pot of 2.2 kg with aluminum reflectors.

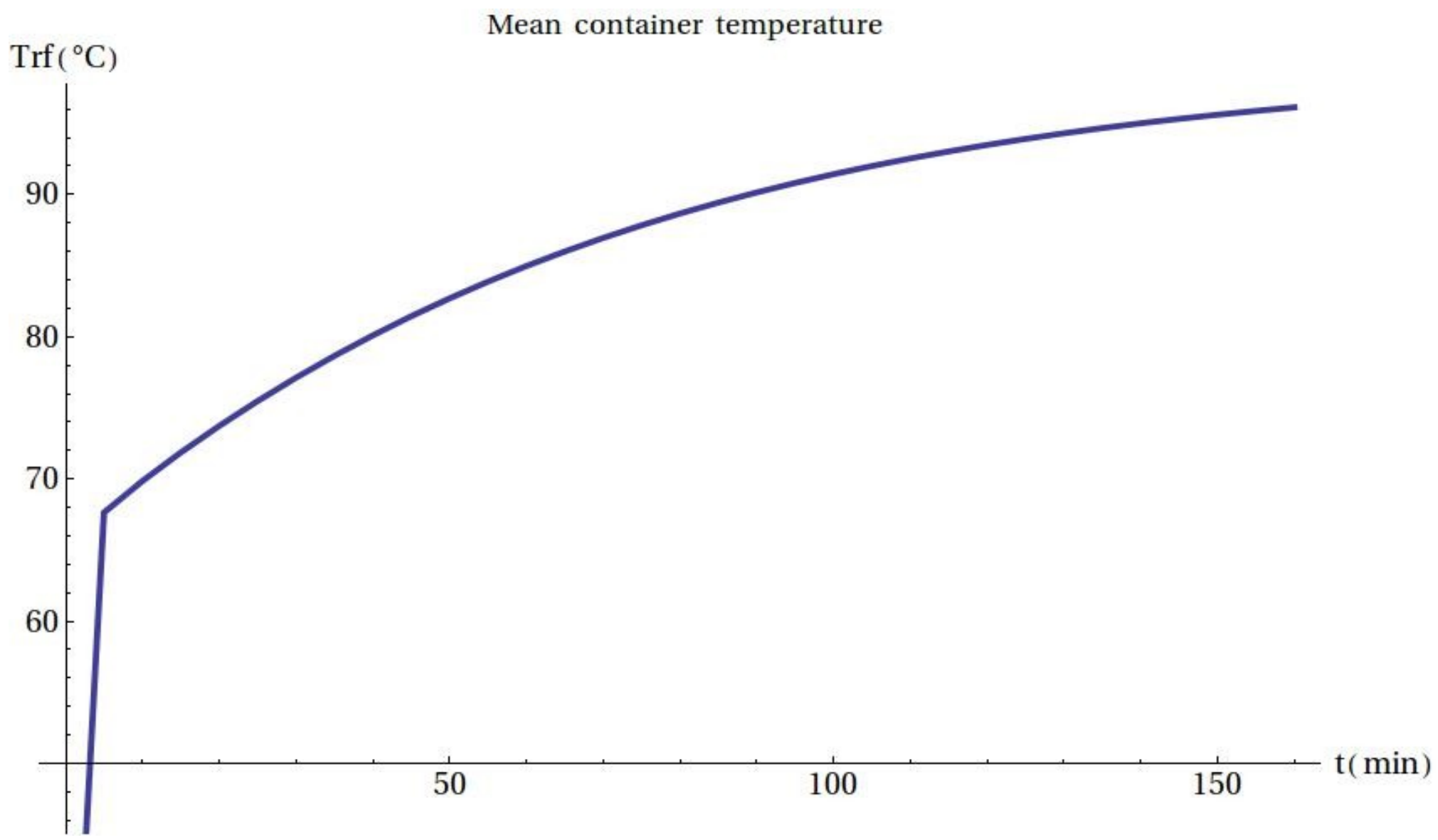

Figure 3.6: Show the mean reflectors temperature versus time corresponding to pressure aluminum pot of 2.2 kg with aluminum reflectors.

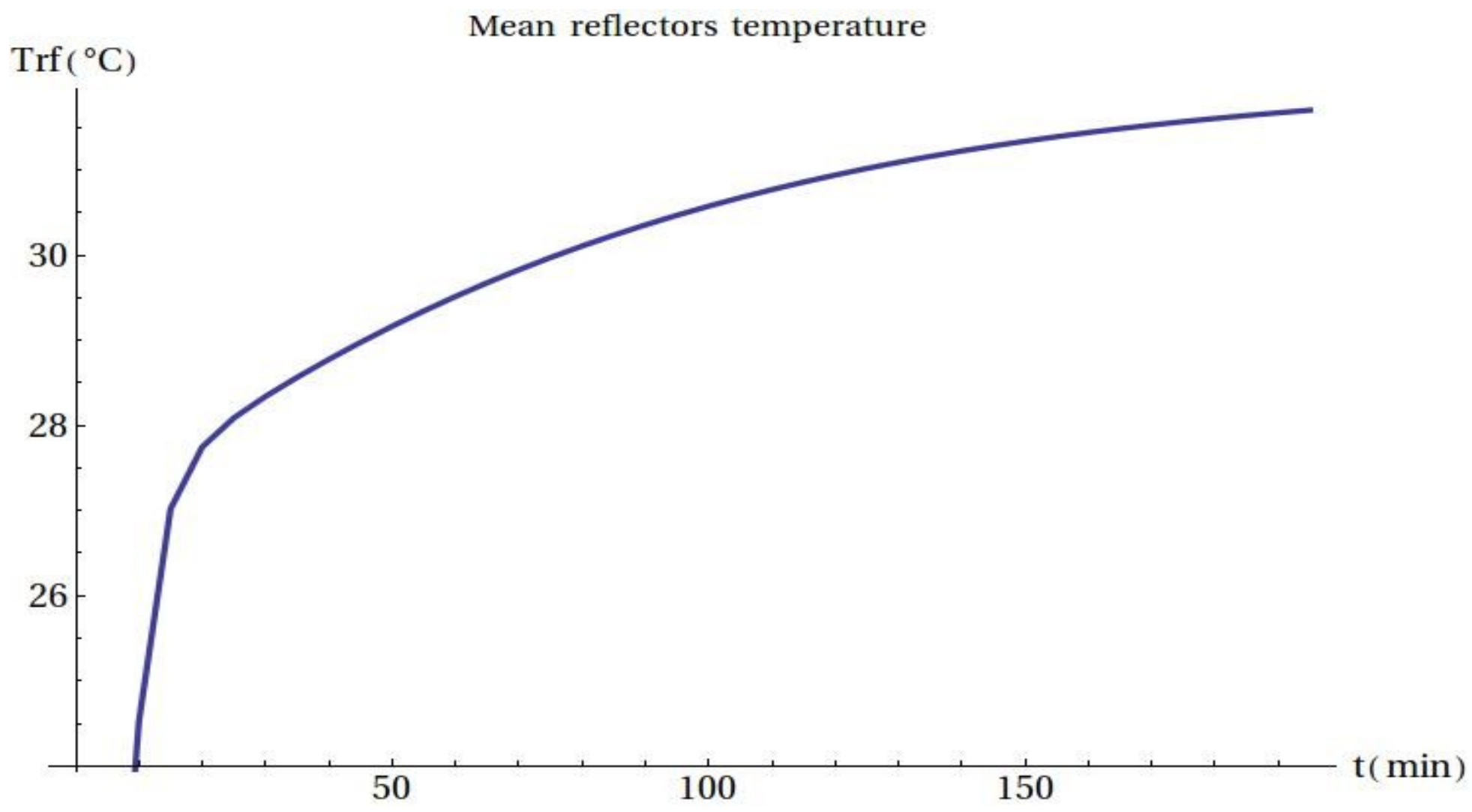

Figure 3.7: Show the mean reflectors temperature versus time corresponding to pressure aluminum pot of 2.2 kg with gift paper as reflectors.

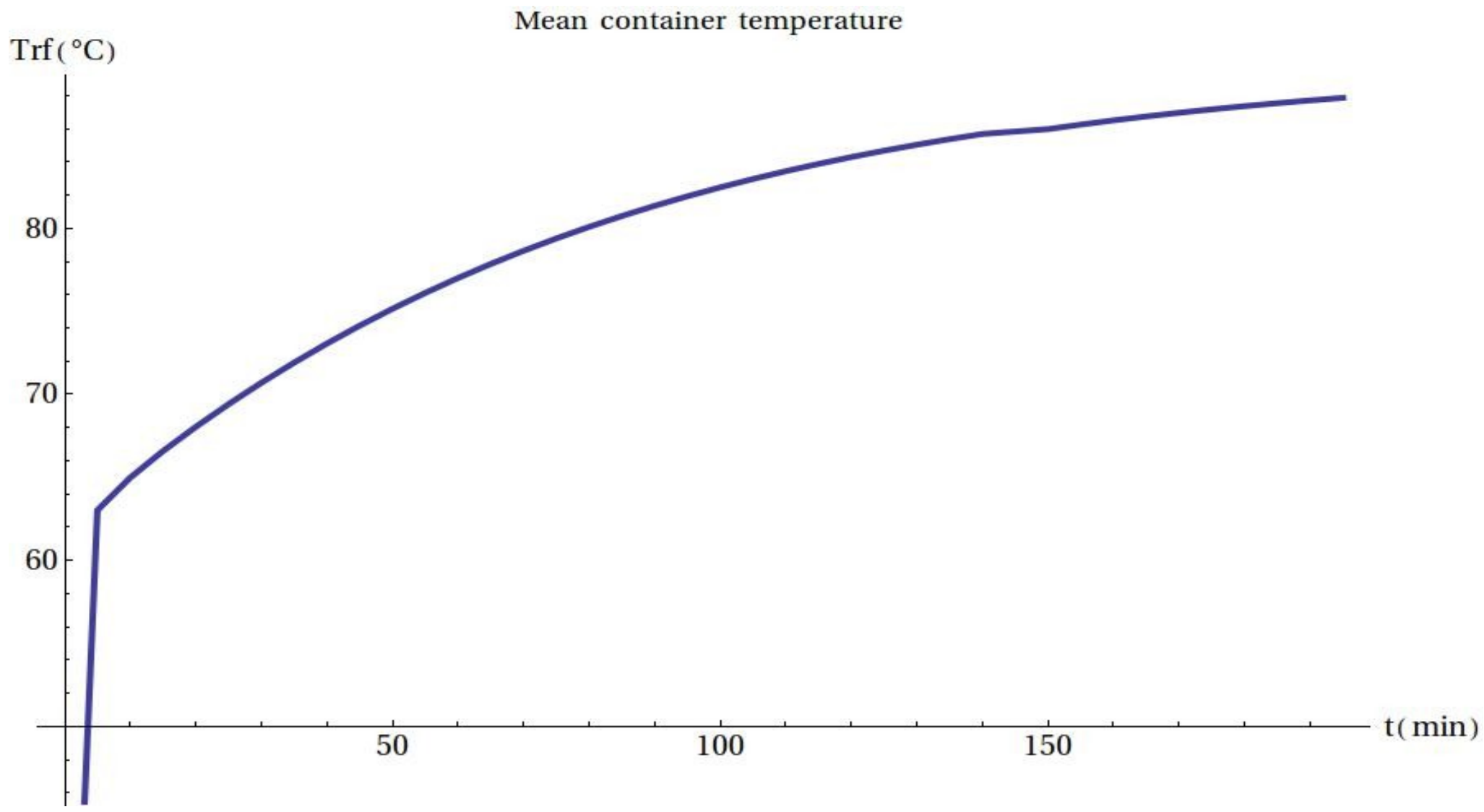

Figure 3.6 Show the mean container temperature versus time corresponding to pressure aluminium pot of 2.2 kg with gift paper as reflectors.

*Tables*

Table 1. It shows the most important results, such as: errors of simulated variables by using our thermal model with respect to experimental data. In the same way it shows standardized cooking power obtained by numerical values with respect to experimental standardized cooking power.

| CASES | RELATIVE | SIMULATED COOKING | EXPERIMENTAL | RELATIVE ERROR OF | SPREAD RELATIVE |
|---|---|---|---|---|---|

| | ERROR (%) | POWER (W) | COOKING POWER (W) | COOKING POWER (%) | ERROR IN EXPERIMENTAL APPROACH (%) |
|---|---|---|---|---|---|
| SIM1 | 2.26 | 115.2 | 123.2 | 6.5 | 14 |
| SIM2 | 6.56 | 104.3 | 118.6 | 12 | 14 |

DISCUSSION

In this paper it has been proposed a thermal model due to simulate thermal behavior of solar cooker “Jorhejpataranskua”, in terms of a system of three coupled nonlinear differential equations. The unknowns of differential equations are: mean water temperature, mean absorber container temperature and mean surface reflectors temperature. It was only compared numerical solution of mean water temperature with respect to experimental data obtained by heat testing of water on solar cooker. It found good agreement between numerical solutions and experimental data with relative errors less than 4%, in many cases. Further experimental data of mean water temperature has an error of +- 0.34 degrees. It was not possible to make comparisons of the other two temperatures because we have not reliable experimental data.

Moreover it has been used numerical results of thermal model to calculate cooking power for two different container pot. So that it was obtained satisfactory results with differences less than 12%, but these results correspond to a good approximation, because of estimation differences are less than propagation errors of experimental testing.

CONCLUSION

The thermal model developed in this paper gives us a good approximation with respect to experimental data in the case of mean water temperature. For mean reflectors temperature and mean container temperature are physically consistent with experimental observation, nevertheless for these cases we do not have reliable experimental data.

This thermal model can be useful to calculate standardized cooking power of solar cookers as “Jorhejpataranskua” without making a device for standard testing. It was possible to calculate standardized cooking power for two cases: pressure aliminum pot and stainless steel pot.

It is also possible to vary parameters such as catchment area of collector and absorber container, optics properties of reflective sheets, selective film of absorber, among others due to find more interesting thermal properties of a solar cooker.

**ACKNOWLEDGEMENTS**

We thank the support of the CONACyT project number 166126 and Universidad Intercultural Indígena de Michoacán.

**Apéndice A.**

```
Sistemasim =
 FullSimplify[sistema = {Tr1'[t] == 1/(mr * cr) * (Q1rad - Q2rad - Q3conv - Q4rad - Q8conv - Q9rad - Q3conv2), Trf'[t] == 1/(mrf * crf) * (Q5rad + Q4rad - Q6conv - Q7rad + Q3conv2),
   Tf'[t] == 1/(mf * cf) * (Q8conv + Q9rad)}]
sol = NDSolve[{Sistemasim, Trf[0] == Tamb, Tr[0] == Tamb, Tf[0] == Tamb}, {Trf, Tf, Tr}, {t, 0, 7000}, Method -> "Automatic"]
```